\documentstyle[proceedings,numreferences,epsfig]{crckapb}

\begin{opening}
\title{TESLA Polarimeters}
\author{V.Gharibyan, N.~Meyners, K.~P.~Sch\"uler}
\institute{DESY,~Deutsches~Elektronen~Synchrotron,~Hamburg,~Germany }
\end{opening}
\runningtitle{TESLA Polarimeters}
\begin{document}
\begin{abstract}
We describe a study of high-energy Compton beam polarimeters for the future
$e^+e^-$ linear collider machine TESLA.
A segment of the beam delivery system has been identified, which is aligned with 
the $e^+e^-$ collision axis and which has a suitable configuration for high-quality
beam polarization measurements. The laser envisaged for the polarimeter is similar
to an existing facility at DESY. It delivers very short pulses in the 
$10~ps,~10-100~\mu J$ regime and operates with a pattern that matches the pulse and bunch
structure of TESLA. This will permit very fast and accurate measurements
and an expeditious tune-up of the spin manipulators at the low-energy end of the linac.
Electron detection in the multi-event regime will be the principle operating mode of
the polarimeter. Other possible operating modes include photon detection and
single-event detection for calibration purposes. We expect an overall precision of
$\Delta P/P \sim 0.5\%$ for the measurement of the beam polarization.
\end{abstract}
\vspace{-12mm}
\section{Introduction and Overview}
A full exploitation of the physics potential of TESLA must aim to employ 
polarized electron and positron beams with a high degree of longitudinal
polarization at full intensity. The technology of polarized electron sources 
of the strained GaAs type is well established~\cite{slacsource,cdr} and 
TESLA is therefore likely to deliver a state of the art polarized electron beam 
with about 80\% polarization from the very beginning.
The development of suitable beam sources of polarized positrons, based on the 
undulator method~\cite{balmik,flot} is still in its infancy but may soon be 
started with real beam tests at SLAC~\cite{e166}. 
Equally important to the generation of high beam polarization will be its 
precise measurement and control over the full range of planned beam energies
(45.6, 250, $400~GeV$).
The quantity of basic interest is the
longitudinal spin polarization of the two beams at the interaction point. 
Since a precise polarization measurement at the detector IP itself is
difficult, the point of measurement should be chosen
such that beam transport and beam-beam interaction effects are either
negligible or small and well quantified. Other important factors relate
to the level of radiation backgrounds and to the technical infrastructure
and accessibility of a chosen site.
\begin{figure}[hbt]
\vspace{5mm}
\centerline{\epsfig{file=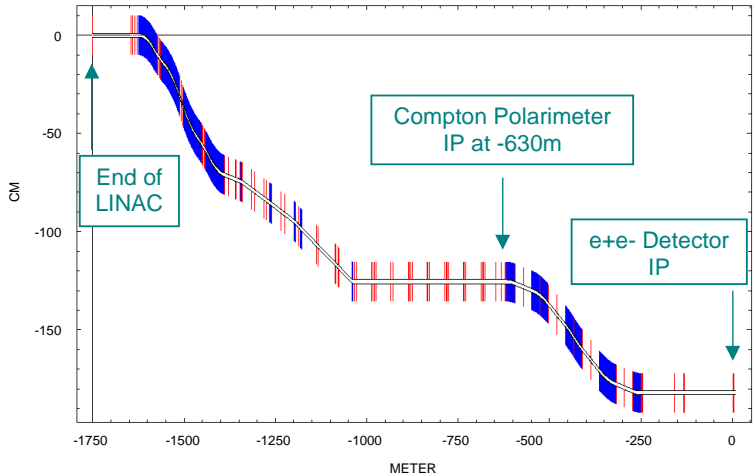, width=10cm, angle=0}}
\caption{TESLA Beam Delivery System}
\end{figure}
The concept of the polarimeter that we propose for TESLA
is based on the well established laser backscattering method, as it was already
envisaged in the TESLA CDR~\cite{cdr}. 
The proposed location of the Compton IP, where the laser beam crosses the
electron or positron beam, is 630 meters upstream of the center of the
$e^+e^-$ detector, near the end of a long straight section of the beam delivery
system (BDS), see Fig.~1.
This part of the beamline is foreseen for general beam diagnosis
and is also well suited for high quality beam polarization measurements.

Although the polarization vector experiences large rotations 
(due to the g-2 effect) as the beam traverses the bends of the BDS, 
the beam and spin directions at the chosen polarimeter site are precisely aligned,
except for a parallel offset, with those at the $e^+e^-$ interaction point.
A polarization measurement at the proposed upstream location will
therefore provide a genuine determination of the quantity of interest,
as long as beam-beam effects are negligible or correctable. This is indeed
the case. We estimate the beam-beam induced depolarization at TESLA to be
0.5\%. 

Fig.~2 shows a layout of the Compton Polarimeter. The laser beam
crosses the electron or positron beam with a small crossing angle of
10~mrad at z~=~-630~m, just upstream of a train of ten C-type dipole 
magnets (BFCH0) which bend the beam horizontally by 0.77~mrad. 
The Compton scattered electrons are momentum analyzed in the field of
the dipoles and detected with a segmented 14 channel gas Cerenkov counter.
An optional calorimetric photon detector can also be employed 
further downstream.
\begin{figure}[hbt]
\centerline{\epsfig{file=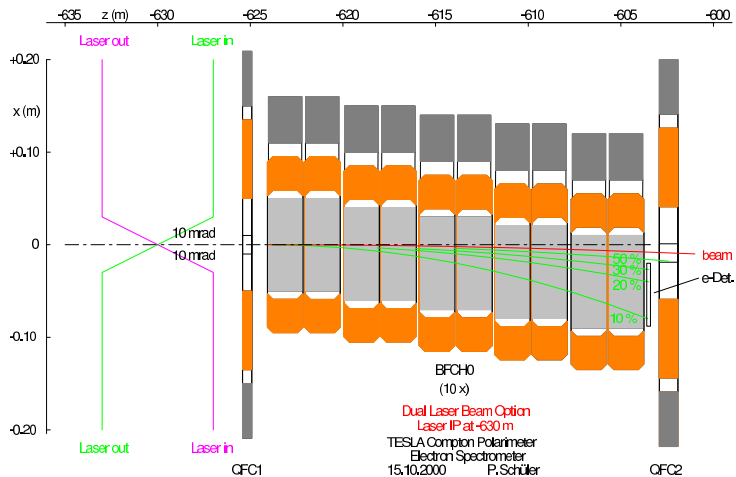,height=10cm,angle=-90}}
\caption{Layout of the Compton Polarimeter}
\end{figure}

The laser system that we envisage for the polarimeter should be similar to the
laser configuration that has been developed by Max Born Institute for the Tesla
Test Facility (TTF) photo injector gun at DESY~\cite{lasref1}.
This laser can be pulsed with a pattern that matches the peculiar pulse
and bunch structure of TESLA. In this way it is possible to achieve very high
luminosity, typically six orders of magnitude higher than with continuous
lasers of comparable average power. 

The statistics of Compton produced events
is very high to the point where statistical errors will not matter in 
comparison with systematic errors. We expect a performance similar to 
the SLD Compton polarimeter at SLAC~\cite{woods}, with an overall precision of
$\Delta P / P \sim 0.5\%$ for the measurement of the beam polarization.

\section{General Considerations}
The spin motion of a deflected electron or positron beam in a transverse 
magnetic field follows from the familiar Thomas-Larmor expression

\begin{equation}
\theta^{spin} \;=\; \gamma\: \frac{g-2}{2}\: \theta^{orbit} 
\;=\; \frac{E_0}{0.44065~GeV}\: \theta^{orbit}
\label{eq:spinmotion} \\
\end{equation}
\noindent
where $\theta^{orbit}$ and $\theta^{spin}$ are the orbit and spin 
deflection angles, $E_0$ is the beam energy, $\gamma=E_0/m$, and 
$(g-2)/2$ is the famous g-factor anomaly of the magnetic moment of
the electron.  

In order to guarantee that the polarization measurement $\Delta P/P$
at the chosen polarimeter site does not suffer from systematic 
misalignments of the beam direction,
we will postulate the following alignment tolerances 
\begin{eqnarray*}
\Delta P/P  \leq  0.1\% \; \longrightarrow \; 
\Delta \theta^{spin} \leq 45\,mrad
& \; \longrightarrow \; & \Delta \theta^{orbit} \leq 50\,\mu rad
\; 
\end{eqnarray*}
where $\theta$ denotes the polar angle.
The beam direction at the polarimeter site should therefore be aligned 
with the collision axis at the $e^+e^-$ interaction point to within 
$50\mu rad$.

The strong beam-beam interaction at the collider IP will diffuse the
angular spread of the beam. In Table~\ref{table:beam-beam} we have 
listed the rms values of the orbital angular spread of the disrupted 
beams at TESLA as obtained by O. Napoly. From the orbital rms values
we have determined the associated rms spin distribution angles
which are listed in Table~\ref{table:beam-beam}. 
Based on these numbers,
we estimate the overall depolarization of the spent beam to be
$\Delta P/P \simeq 1 - cos (139mrad) = 1\%$, independent of beam energy.
Assuming that the beam-beam interaction proceeds in a 
symmetric fashion upstream and downstream from the IP, we estimate
the effective depolarization of the beam before the IP to be half
of the overall effect, i.e. $0.5\%$. 
\begin{table}[htb]
\begin{center}
\begin{tabular}{|c c c c c|} 
\hline
   & $\Delta\theta_{x}^{orbit}(rms)$ & $\Delta\theta_{y}^{orbit}(rms)$
   & $\Delta\theta_{x}^{spin}(rms)$  & $\Delta\theta_{y}^{spin}(rms)$ \\
   & $(\mu rad)$  & $(\mu rad)$    &   $(mrad)$   &   $(mrad)$  \\
\hline
250~GeV &   245        &      27        &    139     &    15     \\
400~GeV &   153        &      17        &    139     &    15     \\
\hline
\end{tabular}
\caption[Beam beam effects]{\label{table:beam-beam}
Disrupted beam rms angular spreads of orbit and spin angles. }
\end{center}
\end{table}

\begin{figure}[h]
\centerline{\epsfig{file=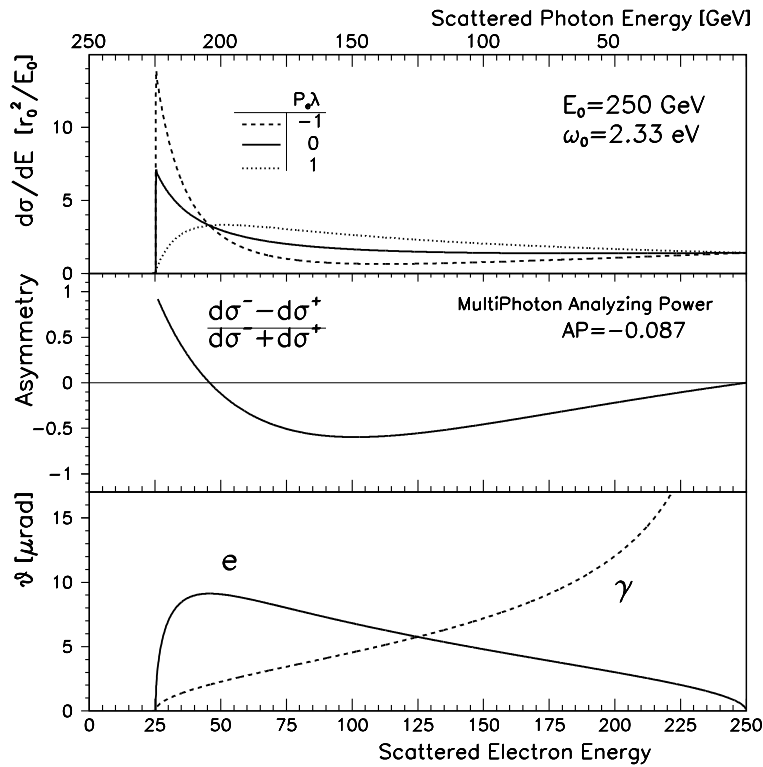, width=10cm, angle=0}}
\caption{Energy spectra (top), spin asymmetry (middle) 
and scattering angles (bottom) of Compton scattered electrons and photons,
for a beam energy of 250~GeV and a green laser.}
\end{figure}

\vspace{-7mm}
\section{Compton Polarimeter}
The Compton kinematics are characterized by the dimensionless variable
\begin{equation}
x =  \frac{4E_0 \omega_0}{m^2} \cos^2{(\theta_0 /2)}
\label{eq:x} \\
\end{equation}
\vspace{0.5cm}
where
$E_0$ is the initial electron energy,
$\omega_0$ is the initial photon energy,
$\theta_0$ is the crossing angle between the electron beam and the laser,
$m $ is the electron mass.

The energy spectra, the associated spin asymmetry and the scattering
angles of the Compton scattered electrons and photons are shown in
Fig.~3 for a beam energy of 250~GeV and a green laser
(2.33~eV).
For much higher or lower beam energies, it will be advantageous to change
the wavelength of the laser. 
The multi-photon analyzing power $A_p$ is also indicated in this figure.

The longitudinal polarization of the electron beam is determined
from the asymmetry of two measurements of Compton scattering with parallel
and antiparallel spin configurations of the interacting electron and laser
beams. 

For the TESLA Compton polarimeter, we plan to employ electron detection in the 
multi-event regime as the principle detection method. We will, however, reserve
the multi-photon detection method as an option, especially for
TESLA operation at the Z-pole.
Furthermore, we would like to point out that it can be very useful for calibration
purposes to operate occasionally in the single-event regime, either with reduced
pulse power of the laser or even with cw lasers.   
\begin{table}[t]
\centerline{\begin{tabular}{|c c c c c c|}\hline
configuration & $E_0$ &$\omega_0$ & $<P_L>$ & $J$ & ${\cal L}$ \\
 & $(GeV)$ &  $(eV)$ & $(W)$ & $(\mu J)$ &
$(10^{32}cm^{-2} \, s^{-1})$ \\
\hline
TESLA-500 & 250 &  2.33 & 0.5 & 35 & 1.5 \\
TESLA-800 & 400 &  1.165& 1.0 & 71 & 6.0 \\
Giga-Z    & 45.6&  4.66 & 0.2 & 14 & 0.2 \\ 
\hline
\end{tabular}}
\caption[Reference Parameters]{Reference parameters for statistical tables.}
\end{table}
\begin{table}[t]
\centerline{\begin{tabular}{|c c c c c c c c c|}\hline
$bin$ & min $x_{d}$ & max $x_{d}$ & $E/E_0$ & $E/E_0$ & 
Analyzing & Stat. &  $<d\sigma/dE>dE$ & Rate \\
\# &  (mm) & (mm) & low & high & Power & Weight & (mbarn) &  (MHz) \\
\hline
 3 & -75 & -80 & 0.100 & 0.107 &  0.927 & 0.355 & 3.35 & 0.503\\ 
 4 & -70 & -75 & 0.107 & 0.115 &  0.812 & 0.297 & 3.79 & 0.568\\ 
 5 & -65 & -70 & 0.115 & 0.123 &  0.687 & 0.232 & 3.92 & 0.588\\ 
 6 & -60 & -65 & 0.123 & 0.134 &  0.554 & 0.165 & 4.14 & 0.621\\ 
 7 & -55 & -60 & 0.134 & 0.146 &  0.415 & 0.099 & 4.37 & 0.655\\ 
 8 & -50 & -55 & 0.146 & 0.161 &  0.268 & 0.044 & 4.70 & 0.705\\ 
 9 & -45 & -50 & 0.161 & 0.178 &  0.114 & 0.008 & 5.10 & 0.765\\ 
10 & -40 & -45 & 0.178 & 0.201 & -0.044 & 0.001 & 5.57 & 0.835\\ 
11 & -35 & -40 & 0.201 & 0.229 & -0.203 & 0.026 & 6.28 & 0.943\\ 
12 & -30 & -35 & 0.229 & 0.268 & -0.355 & 0.075 & 7.25 & 1.087\\ 
13 & -25 & -30 & 0.268 & 0.321 & -0.489 & 0.133 & 8.74 & 1.311\\ 
14 & -20 & -25 & 0.321 & 0.401 & -0.577 & 0.176 &11.28 & 1.692\\ 
\hline
all& -20 & -90 & 0.089 & 0.401 &        &       &68.49 &10.273\\
\hline
\multicolumn{9}{|c|}{Statistical Error for $\Delta t \: = \:$ 1~second:
$\; \; \; \; \;  \Delta P/P \: = \: 0.89 \cdot 10^{-3}$ } \\
\hline
\end{tabular}}
\caption[Statistical Table 1]{\label{table:stat_1}
Event rates and statistical error for TESLA-500.}
\end{table}
For the determination of event rates and statistical errors,
we will use the reference parameters listed in Table~2, where $P_L$
($J$) is the laser average(pulse) intensity and ${\cal L}$ is luminosity.
In order to be consistent with the cross sections in Fig.~3, we list 
the wavelengths for a Nd:YAG laser. The wavelengths of the Nd:YLF laser
are only slightly different. Not explicitly listed are the crossing angle
${\theta}_0 \, = \, 10~mrad$ and the size of the laser focus 
${\sigma}_{x\gamma} \, = \, {\sigma}_{y\gamma} \, = \, 50~\mu m$, which
are assumed to be common for all configurations.

The Table~\ref{table:stat_1}
lists the binned cross sections and event rates in the electron
detector ($E$ and $x_d$ are the scattered electron energy and position)
for the first reference configuration of Table~2.
As these events are bunched and recorded as analog signals at the bunch
crossing frequency, there is no problem with apparently high rates,
as we do not actually count individual events.

\begin{table}[t]
\centerline{\begin{tabular}{|l c c|}\hline
                           & $e^+/e^-~beam$     & laser~beam          \\ \hline
energy                     & 250~GeV            & 2.3~eV              \\
charge or energy/bunch     & $2 \cdot 10^{10}$  & $35~\mu J$          \\
bunches/sec                & 14100              & 14100               \\
bunch length $\sigma_t$    & 1.3~ps             & 10~ps               \\
average current(power)     & $45~\mu A$         & 0.5~W               \\
$\sigma_x \cdot \sigma_y~(\mu m)$ & $10 \cdot 1$   & $50 \cdot 50$    \\ \hline
beam crossing angle        & \multicolumn{2}{c|}{10~mrad}            \\
luminosity   & \multicolumn{2}{c|}{$1.5 \cdot 10^{32}cm^{-2}s^{-1}$} \\
cross section& \multicolumn{2}{c|}{$0.136 \cdot 10^{-24}cm^2$}       \\
detected events/sec        & \multicolumn{2}{c|}{$1.0 \cdot 10^7$}   \\
detected events/bunch      & \multicolumn{2}{c|}{$0.7 \cdot 10^3$}   \\
$\Delta P / P$ stat. error/sec & \multicolumn{2}{c|}{negligible}     \\
$\Delta P / P$ syst. error     & \multicolumn{2}{c|}{$\sim 0.5\%$}   \\ \hline
\end{tabular}}
\caption[Compton Polarimeter Parameters]{\label{table:par250}Compton Polarimeter
Parameters at 250~GeV}
\end{table}
\begin{figure}[h]
\centerline{\epsfig{file=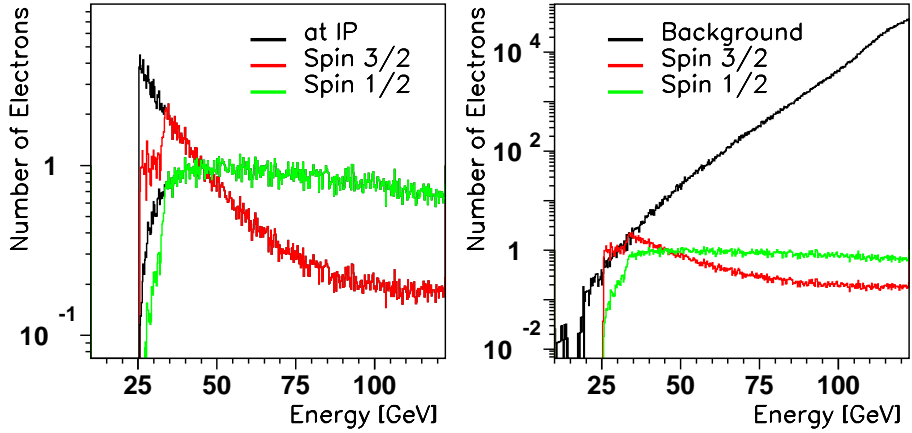, width=12cm, angle=0}}
\caption{Simulated per bunch spectra of scattered electrons for two laser helicities 
tracked down to $z=65~m$. Superimposed: spectra at $z=0.8~m$ (left) and background
at $z=65~m$ (right).}
\end{figure}
The analyzing power $A_i$ for each bin and the associated statistical
weights $w_i$ for a beam polarization P~=~0.80 are also given in
Table~\ref{table:stat_1}. Furthermore, we list the 
statistical errors $\Delta P/P$ of the beam polarization
for a measurement duration $\Delta t$ of 1~second. 
We conclude from these numbers that genuine statistical errors originating
from the Compton event statistics will be exceedingly small and likely
negligible in comparison with systematic effects.
We expect an overall precision of $\Delta P / P \sim 0.5\%$ for the measurement of
the beam polarization. As an example, Table~\ref{table:par250} gives typical
polarimeter parameters for TESLA-500. The performance is similar for other 
energy regimes of TESLA.
More details are given in~\cite{lcnot}. 

We have also considered the possibility of downstream polarimeter
locations, which would in principle permit to investigate beam-beam
effects experimentally.
However, disrupted beam polarimetry appears very difficult because of 
severe background from energy-degraded beam electrons as shown in Figure 3 
for $250~GeV$ electrons, green laser and nominal beam extraction parameters 
with reduced $10~mm$ collimation at $z=18~m$.

\end{document}